# Strong similarities between the local electronic structure of insulating iron pnictide and lightly doped cuprate

Cun Ye[1], Wei Ruan[1], Peng Cai[1], Xintong Li[1], Aifeng Wang[2], Xianhui Chen[2], and Yayu Wang[1,3,†]

[1]*State Key Laboratory of Low Dimensional Quantum Physics, Department of Physics, Tsinghua University, Beijing 100084, P.R. China*

[2]*Hefei National Laboratory for Physical Science at Microscale and Department of Physics, University of Science and Technology of China, Hefei, Anhui 230026, P.R. China*

[3]*Collaborative Innovation Center of Quantum Matter, Beijing 100084, P.R. China*

[†]Email: yayuwang@tsinghua.edu.cn

One of the major puzzles regarding unconventional superconductivity is how some of the most interesting superconductors are related to an insulating phase that lies in close proximity. Here we report scanning tunneling microscopy studies of the local electronic structure of Cu doped NaFeAs across the superconductor to insulator transition. We find that in the highly insulating regime the electronic spectrum develops an energy gap with diminishing density of state at the Fermi level. The overall lineshape and strong spatial variations of the spectra are strikingly similar to that of lightly doped cuprates close to the parent Mott insulator. We propose that the suppression of itinerant electron state and strong impurity potential induced by Cu dopants lead to this insulating iron pnictide.

A key task in unraveling the mystery of unconventional superconductivity is to disentangle the various phases that lie in close proximity to the superconducting (SC) phase. One of the most surprising discoveries in this regard is that the cuprate superconductors, which possess the highest transition temperature ($T_C$), have an insulating parent compound. How the high $T_C$ superconductivity emerges from doping charges into a Mott insulator has since become one of the greatest challenges in modern condensed matter physics [1]. Although the answer remains controversial, there is general consensus that the Mott insulator represents a valid starting point for tackling the problem. The recently discovered iron-based superconductors, on the contrary, have a metallic parent state with well-defined Fermi surface [2, 3]. This apparently implies a totally different viewpoint for understanding the mechanism of superconductivity. Most theoretical proposals start from the itinerant electron picture and search for the excitations that mediate the Cooper pairs [4, 5]. The electron correlation effect, which plays a dominant role in the cuprates, is generally considered to be unimportant in the iron pnictides.

An intriguing question inspired by this sharp contrast is whether there also exists an insulating phase in the iron pnictides where the electron correlation effect is essential. If the answer is yes, then this insulating phase may be the missing piece of the puzzle that links the pnictides to the cuprates. However, in most iron pnictides the effect of transition metal doping is to introduce charge carriers, suppress the spin density wave order, and induce superconductivity [6, 7]. Overdoping usually leads to a non-SC phase with better metallicity, and an insulating phase is very rare. Interestingly, it was found recently that Cu doped NaFeAs exhibits a superconductor to insulator transition (SIT) with increasing Cu content [8]. This system thus provides a unique opportunity to explore the insulating regime of iron pnictides that was not available before. Understanding the electronic structure of the insulating ground state could shed new lights on the mechanism of iron-based superconductivity and its connection to the cuprates.

In this work we use scanning tunneling microscopy (STM) to investigate the local electronic structure of NaFe$_{1-x}$Cu$_x$As with varied Cu contents across the SIT. We find that Cu dopants induce a systematic depletion of the itinerant electron spectral weight. In the highly insulating regime the electronic spectra develop an energy gap with diminishing density of state (DOS) at the Fermi level ($E_F$), strikingly similar to that in lightly doped cuprates close to the parent Mott insulator. We propose that the suppression of itinerant electron state and strong impurity potential induced by Cu dopants lead to the insulating phase, in which electron correlation effect might be relevant.

The 111-type NaFeAs is ideal for STM studies due to the charge neutral cleavage surface between two neighboring Na layers, on which the electronic structure reflects the intrinsic property of the bulk [9]. Figure 1(a) inset depicts the crystal structure of NaFeAs doped with Cu impurity. Cu substitution of Fe is expected to create a rectangular shaped pattern on the surface Na layer, as marked by the gray plane. The main panel of Fig. 1(a) illustrates the schematic phase diagram of NaFe$_{1-x}$Cu$_x$As, which has a substantially suppressed SC dome than the widely studied Co doped NaFeAs. A more fundamental difference is at highly Cu doped regime an insulating phase emerges, whereas increasing Co content makes the system more metallic [10]. The three solid symbols along the $x$-axis in Fig. 1(a) indicate the three doping levels covered in this work, and Fig. 1(b) displays their normalized resistivity ($\rho$) vs. temperature ($T$) curves. The $x = 0.02$ sample is optimally doped with $T_C = 11$ K, but the $x = 0.14$ sample shows weak insulating behavior at low $T$. The $x = 0.30$ sample represents the highest Cu doping available so far and shows strongly insulating behavior. Fig. 1(c) shows the $T$ dependence of magnetic susceptibility up to 400 K. The general trend is a deviation from the high $T$ linear behavior [11] with increasing Cu content, and at $x = 0.30$ a pronounced Curie-Weiss-like upturn appears below $T \sim 200$ K.

We first use STM to examine the optimally doped sample with dilute Cu dopants. Shown in Fig. 2(a) is a topographic image acquired on a cleaved surface of the $x = 0.02$ sample, where the square lattice with lattice constant $a = 4$ Å is clearly resolved. The Cu dopants can be identified as the bright dumbbells centered on the Fe sites as marked by the dashed boxes, which is consistent with the expected rectangle shown in Fig. 1(a) inset. Fig. 2(b) displays spatially-averaged differential conductance ($dI/dV$) taken at locations void of impurities. At $T = 5$ K, the spectrum shows a well-defined SC gap with symmetric coherence peaks on both sides of $E_F$. The gap amplitude estimated from the separation between the two peaks is $2\Delta = 9$ meV, comparable to the 11 meV gap in optimally doped $NaFe_{1-x}Co_xAs$ [12] with $T_C = 20$ K. With increasing $T$ the gap gradually fills up and the sharp coherence peaks disappear above $T_C = 11$ K.

To scrutinize how Cu dopant affects the superconductivity, we show in Fig. 2(c) a series of $dI/dV$ spectra taken in the vicinity of a Cu dopant (inset). As we move towards the impurity center, the main effect is the suppression (enhancement) of coherence peak on the negative (positive) bias side, whereas the gap amplitude remains the same. These features are similar to that reported on Cu doped $NaFe_{1-x}Co_xAs$ [12]. Fig. 2(d) shows the $dI/dV$ curves taken on the same locations but over a larger bias range. The overall lineshape is similar to that in Co doped NaFeAs, showing a peak at negative bias around -200 mV and a DOS minimum right at $E_F$. The most prominent effect of Cu dopant here is to cause a suppression of the -200 mV peak, especially for the two spots close to the impurity site. Such local spectral weight suppression is absent in the Co doped case.

We next move to the weakly insulating $x = 0.14$ sample. The topography shown in Fig. 3(a) becomes much more disordered because of the higher level of Cu doping, making it difficult to resolve the atomic lattice. Fig. 3(b) displays 9 representative *dI/dV* curves measured on various locations marked by the colored dots. The -200 mV peak is continuously suppressed, presumably due to increased Cu content if we assume that the effect of Cu dopant is the same as that shown in Fig. 3(d). On the positive bias side, however, the unoccupied states are almost identical across the whole sample. Shown in Fig. 3(c) are low bias *dI/dV* curves measured on the same locations, which reveal a peak near -20 mV and a dip around $E_F$. More interestingly, the low energy electronic state is closely correlated with the height of the -200 mV peak. The spatial distribution of DOS can be directly visualized by *dI/dV* maps taken at varied biases. As shown in Fig. 4(a), the *dI/dV* map taken at $V = -200$ mV exhibits strong spatial separation into patches with typical size around 2 nm. The *dI/dV* map taken at $V = -15$ mV (Fig. 4(b)) shows similar spatial variations and positive correlation with that in Fig. 4(a). Therefore, the peak-dip feature near $E_F$ is more pronounced in areas with large DOS peak around -200 mV. In contrast, the *dI/dV* map taken at $V = +200$ mV (Fig. 4(c)) is much more homogeneous, indicating the insensitivity of unoccupied states to Cu doping.

Finally we reach the strongly insulating $x = 0.30$ sample. As shown in Fig. 5(a), the topography becomes even more disordered with very small granular size down to 1 nm scale. The large bias *dI/dV* curves shown in Fig. 5(b) exhibit a further suppression of the occupied state spectral weight compared to that of the $x = 0.14$ sample. More interestingly, a closer examination of the low energy electronic state reveals some fundamentally new features. Fig. 5(c) shows that the spectra around $E_F$ develop a small energy gap, and for some of the curves (e.g. the black one) the DOS at $E_F$ is truly zero. The vanishing Fermi surface in this heavily doped regime is apparently consistent with the strongly insulating resistivity behavior. Moreover, there is also a close correlation between the high energy and low energy electronic states. The areas with strongly suppressed occupied state spectra, presumably with higher Cu content, possess a larger gap size (up to 20 meV) and smaller residual DOS at $E_F$. These features can be directly visualized by the *dI/dV* maps measured at varied bias voltages. As shown in Fig. 6(a) and (b), the locations with strongly suppressed DOS at -200 mV have much reduced residual DOS at $E_F$, as well as a larger energy gap as illustrated by the gap map displayed in Fig. 6(c). The correlation relations between the various quantities are summarized in Fig. 6(d).

A more remarkable feature can be revealed by comparing the spectra of the $x = 0.30$ NaFe$_{1-x}$Cu$_x$As with that of lightly doped cuprates close to the Mott insulator limit. Fig. 5(d) shows the *dI/dV* curves taken on various locations of Na doped $Ca_2CuO_2Cl_2$ with small hole concentrations [13]. The overall spectral patterns of the insulating iron pnictide and lightly doped cuprate are nearly mirror-symmetric when they are put side by side. Over a large bias range, both systems exhibit strongly asymmetric spectral lineshape, where one side shows a steep increase and the other side with lower spectral weight shows a peak. At low bias, both systems develop a small energy gap with the bottom pinned right at $E_F$. In both systems the size of the gap and the residual DOS at $E_F$ show systematic evolution with local doping level as manifested by the large bias spectra.

The local electronic structure of insulating NaFe$_{1-x}$Cu$_x$As and its striking similarity to lightly doped cuprates are remarkable features that have never been observed before in any iron pnictides. The key issue here is the origin of the insulating phase, and the first question to tackle is the effect of Cu doping on the electronic structure of NaFeAs. This is a highly nontrivial task given the fact that for most iron pnictides the low energy electronic structure originates from five partially filled Fe 3$d$ orbitals lying near $E_F$. Some of the orbitals are itinerant, hence are responsible for the Fermi surface formation and metallicity [1]. The other orbitals are more localized, which provide the source for local magnetic moments and spin dynamics [2]. The coexistence and coupling between the local moments and itinerant electrons are responsible for much of the complexities and rich electronic states in the iron pnictides [14-16].

Although there are still controversies regarding the charge and spin state of Cu dopant in iron pnictides [17-20], our $dI/dV$ spectroscopy clear reveals that the main effect of Cu doping is to cause a suppression of the electron DOS on the occupied side, especially near -200 meV. Similar spectral weight suppression has been observed in recent angle-resolved photoemission spectroscopy (ARPES) measurements on NaFe$_{1-x}$Cu$_x$As with $x$ up to 0.14. Moreover, core level spectrum reveals an increase of high energy electronic state at binding energy of -4 eV [19]. Therefore, Cu substitution of Fe selectively reduces the spectral weight of the itinerant orbitals, and the extra electrons are injected into the localized orbitals. This is also consistent with the pronounced Curie-Weiss-type susceptibility due to enhanced local moments in the $x = 0.30$ sample (Fig. 1(c)). With reduced itinerant electron spectral weight and increased local moments, heavily Cu doped NaFeAs is close to the orbital-selected Mott transition [21-24]. The electron correlation effect is expected to be strongly enhanced in this regime [25].

We note that the majority of iron pnictides show weak correlation effect and remain metallic across the whole phase diagram. In this sense the heavily Cu doped NaFeAs is an exception rather than the norm. The reason why this particular system possesses an insulating ground state can be understood from the two unique features of Cu substitution of Fe. The first is the systematic depletion of the itinerant electron spectral weight, which leads to the asymmetric lineshape over the large energy range. In contrast, in Co doped NaFeAs the peak at -200 mV is nearly unaffected by Co [9]. Secondly, Cu dopant induces strong local impurity potential as manifested by the altered coherence peaks in the SC sample and pronounced inhomogeneity in the insulating sample. First principle calculations predicted that in the presence of substitutes with strong impurity potentials, the actual number of coherent carrier will diminish and localized states away from $E_F$ will emerge [26]. Both features are confirmed by our STM data and the ARPES results. The reduced itinerancy and strong disorderness can further enhance the correlation effect because the Coulomb repulsion cannot be efficiently screened under such circumstances [27]. The strong electron-electron interaction may cause a suppression of electron DOS at $E_F$, which naturally explains the small energy gap observed here in heavily Cu doped NaFeAs. In contrast, the low energy spectrum in heavily (11%) Co doped NaFeAs becomes rather featureless, which is characteristic of weakly interacting metal [9].

In summary, STM studies on Cu doped NaFeAs reveal a systematic evolution of the electronic structure from a superconductor to an insulator with spectral features strikingly similar to lightly doped cuprates. The insulating phase is most likely caused by the combined effect of reduced itinerant spectral weight and strong impurity potential due to Cu doping. These results suggest the relevance of correlation effect in Cu doped NaFeAs, which may serves as a link between the iron pnictides and cuprates. However, given the complexity associated with the *dI/dV* spectra of both the cuprates and iron pnictides, we cannot exclude other explanations for the similarities reported here. Other experimental probes, especially resonant inelastic x-ray scattering spectroscopy, are needed to confirm whether the correlation effect plays an essential role in the electronic structure of this insulating iron pnictide.

We thank Z.Y. Weng, G.M. Zhang and Z. Sun for helpful discussions. We thank Y. Kohsaka for the permission of using Fig. 5(d) from Ref. [13]. This work was supported by the

National Natural Science Foundation and MOST of China (2011CBA00101, 2012CB922002, 2015CB921000). AFW and XHC acknowledge support from the 'Strategic Priority Research Program' of Chinese Academy of Sciences (grant No. XDB04040100).

**Figure captions:**

**Figure 1 (a)** Schematic phase diagram of $NaFe_{1-x}Cu_xAs$. (inset) The crystal structure of $NaFe_{1-x}Cu_xAs$. The gray plane on the top corresponds to the exposed Na surface layer. **(b and c)** Temperature dependence of the in-plane resistivity normalized by the room temperature value and the magnetic susceptibility for the three $NaFe_{1-x}Cu_xAs$ samples.

**Figure 2 (a)** STM topography on the $x = 0.02$ SC sample acquired with bias voltage $V = 10$ mV and tunneling current $I = 300$ pA. **(b)** Variable temperature $dI/dV$ spectra (vertically offset for clarity) collected from a defect-free area across $T_C = 11$ K, set up with $V = -22$ mV and $I = 100$ pA. **(c)** Upper panel: $dI/dV$ curves (vertically shifted) at $T = 5$ K taken at various locations (labeled by corresponding colored dot in the inset) in the vicinity of an isolated Cu dopant. The setup conditions are $V = -50$ mV and $I = 100$ pA. Lower panel: spectroscopic variation around the Cu dopant, obtained by subtracting the $dI/dV$ curves in (c) with the one taken 4 lattice away (black). The colors are the same as in (c). **(d)** Large bias $dI/dV$ spectra taken at the same spots as in (c) with $V = -500$ mV and $I = 100$ pA.

**Figure 3 (a)** STM topography on the $x = 0.14$ sample acquired with $V = -50$ mV and $I = 40$ pA. **(b)** $dI/dV$ curves taken at locations marked by the colored dots in (a) with set up conditions $V = 400$ mV and $I = 180$ pA. A close-up look into energy range around $E_F$ is displayed in **(c)**.

**Figure 4 (a, b, c)** $dI/dV$ maps obtained in the same field of view as in Figure 3 (a) with bias voltages -200 mV, -15 mV and 200 mV respectively. (d) Cross correlation plot of the -200 mV (a) and -15 mV (b) $dI/dV$ maps, demonstrating a clear positive correlation.

**Figure 5** **(a)** STM topography on the strongly insulating $x = 0.30$ sample acquired with $V = 100$ mV and $I = 10$ pA. **(b)** Typical *dI/dV* curves (shifted vertically) taken at locations marked by the colored dots in (a), set up accordingly as Fig. 3(b) with $V = 400$ mV and $I = 180$ pA. **(c)** Same as (b) but zoom in on the low energy range near $E_F$, which reveals a V-shaped gap. **(d)** *dI/dV* spectra taken at various locations in lightly hole-doped Mott insulator $Ca_{2-x}Na_xCu_2OCl_2$ ($0.06 \leq x \leq 0.12$). Adapted by permission from Macmillan Publishers Ltd: Nature Physics [13].

**Figure 6 (a, b)** *dI/dV* image over the same field of view as in Fig. 5(a) with bias voltages of -200 mV and 0 mV. **(c)** Spatial distribution of the size of the V-shaped gap shown in Fig. 5(c). **(d)** Cross correlation plot of the -200 mV peak height versus the local insulating gap magnitude (black) and residue DOS at $E_F$ (blue), revealing negative and positive correlations respectively.

Figure 1

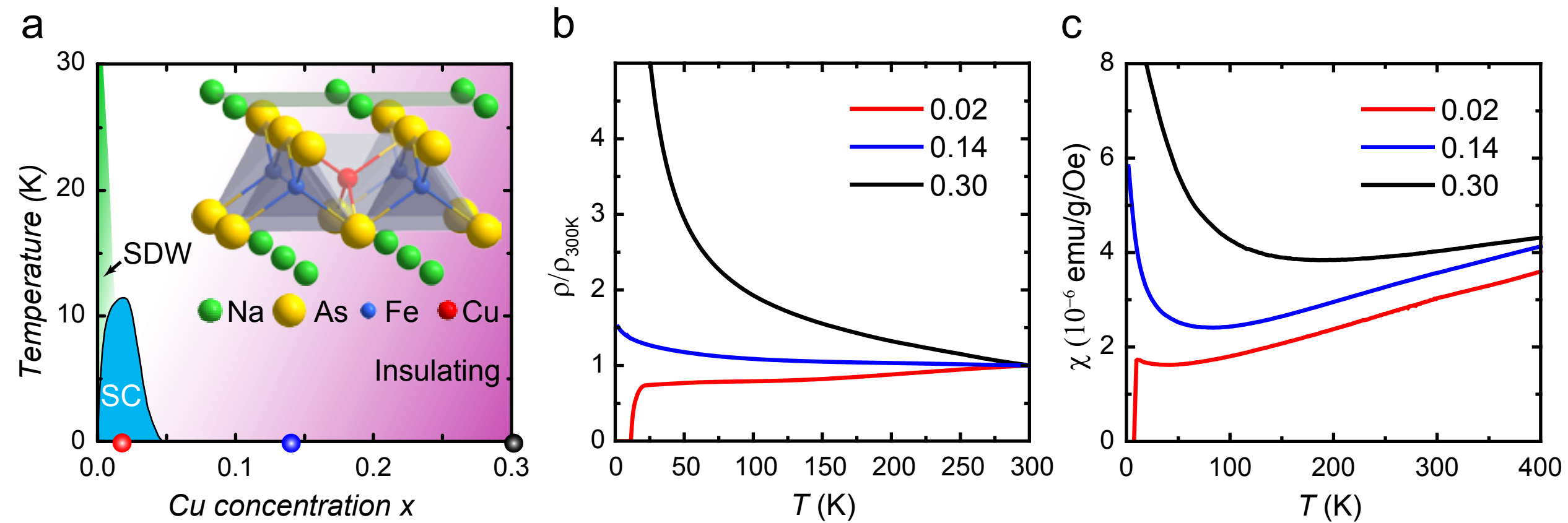

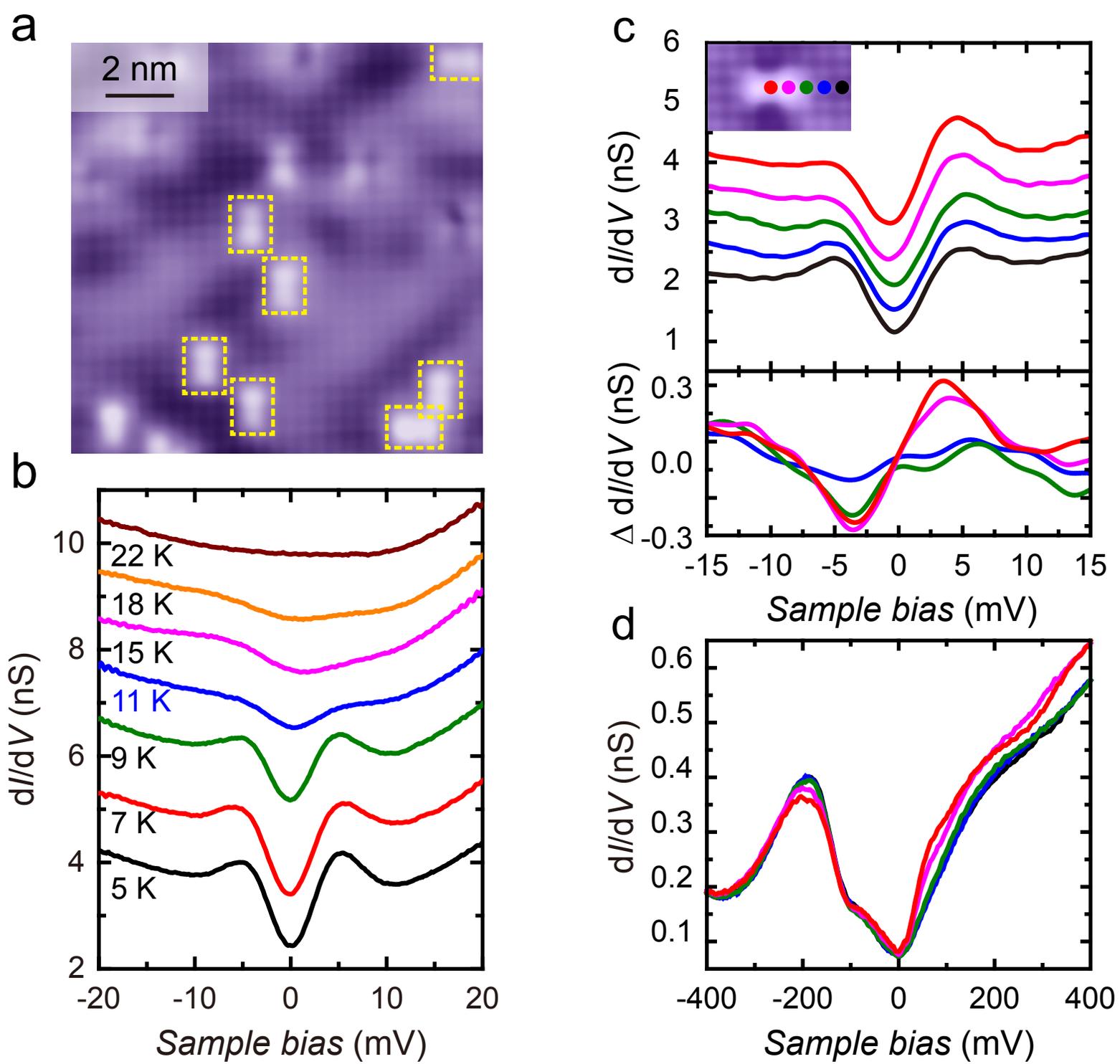

Figure 2

Figure 3

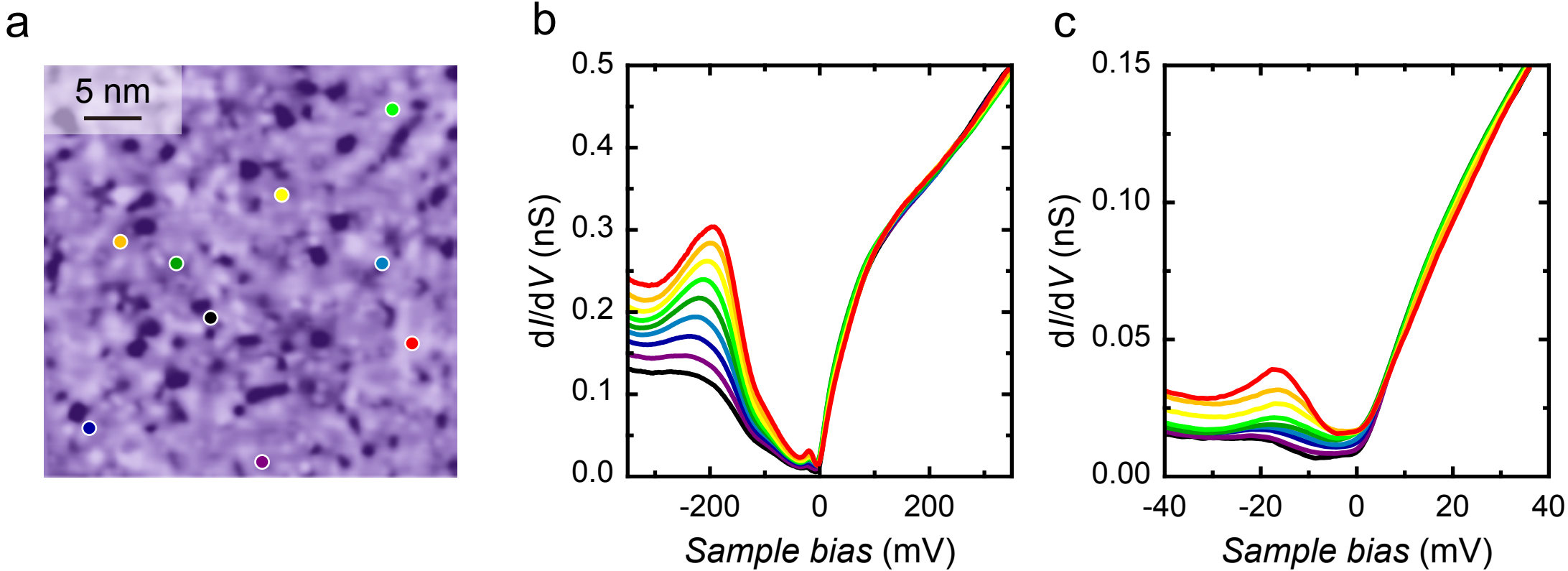

Figure 4

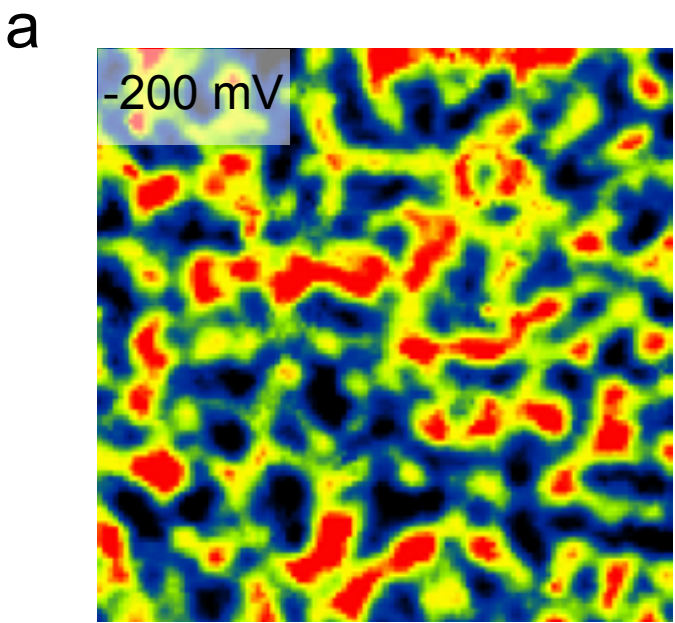
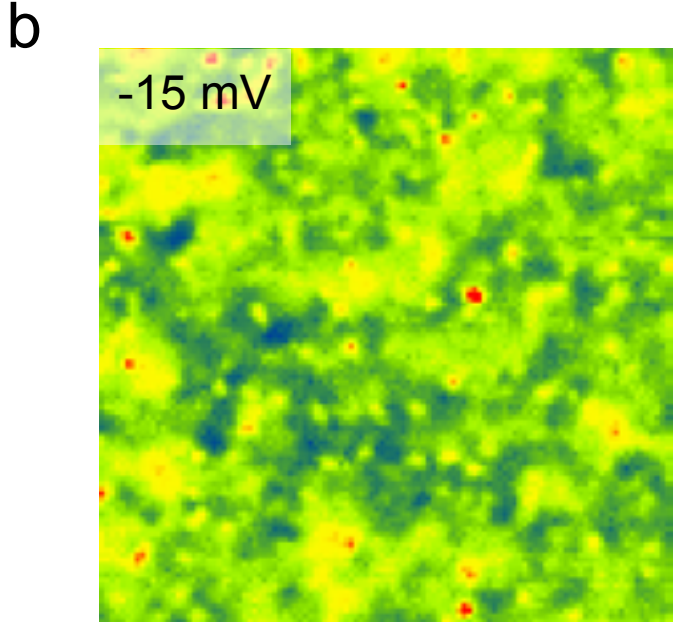
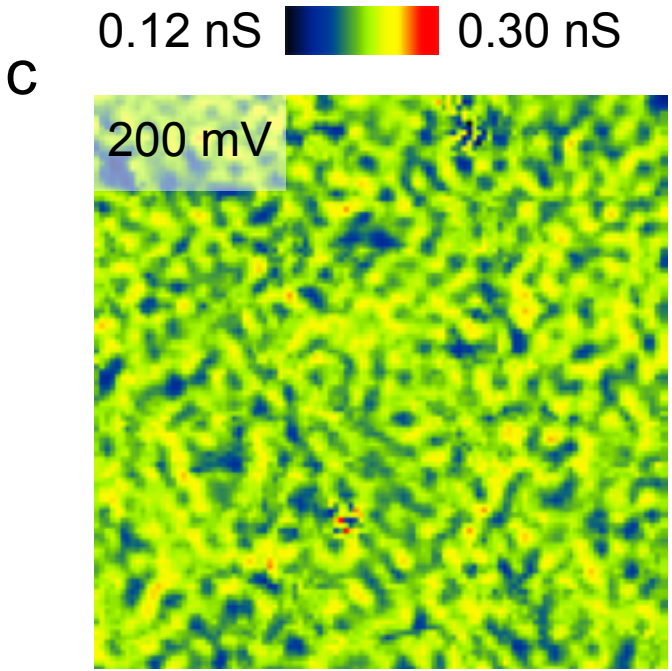
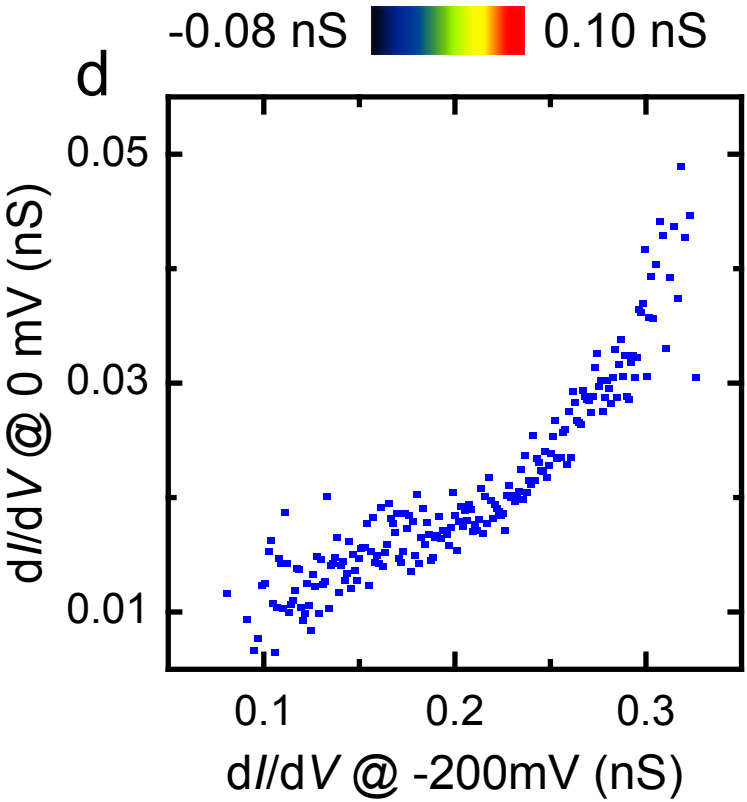

Figure 5

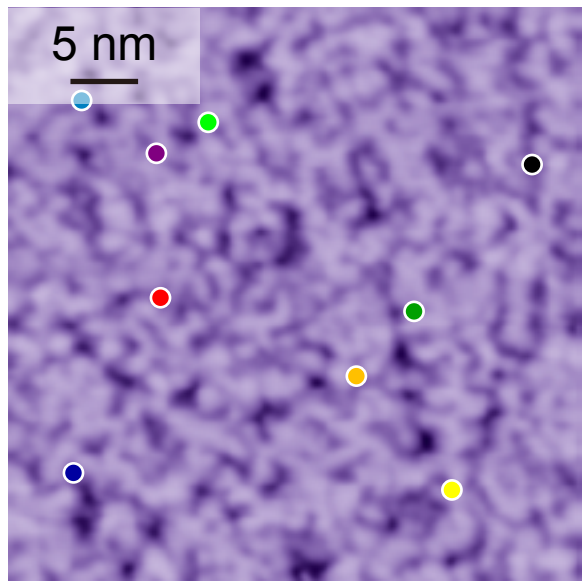
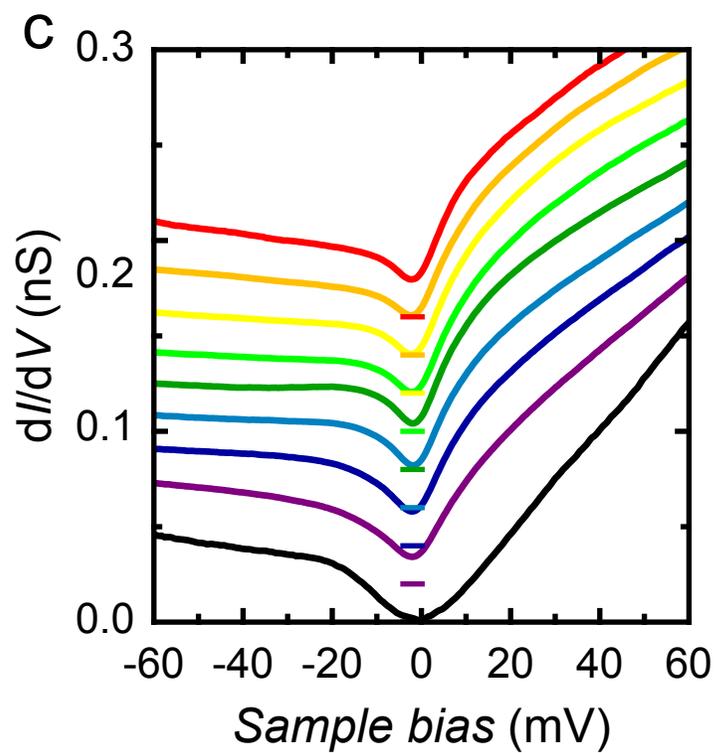
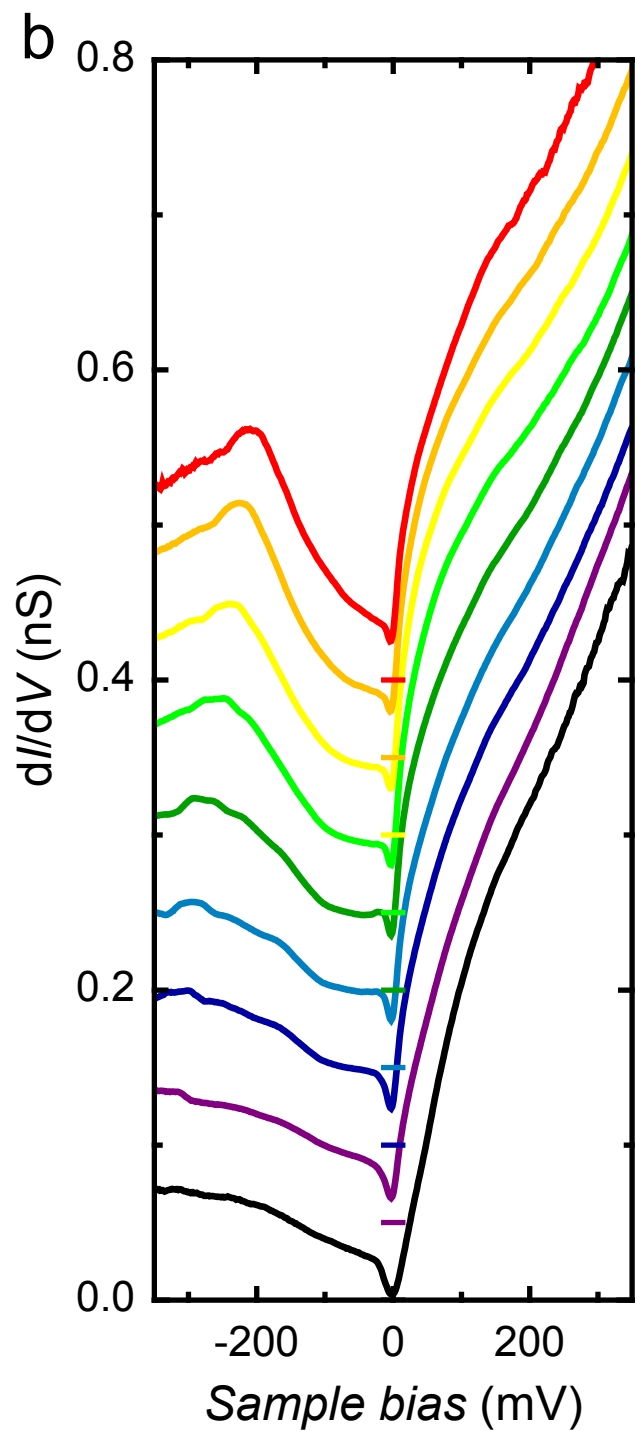
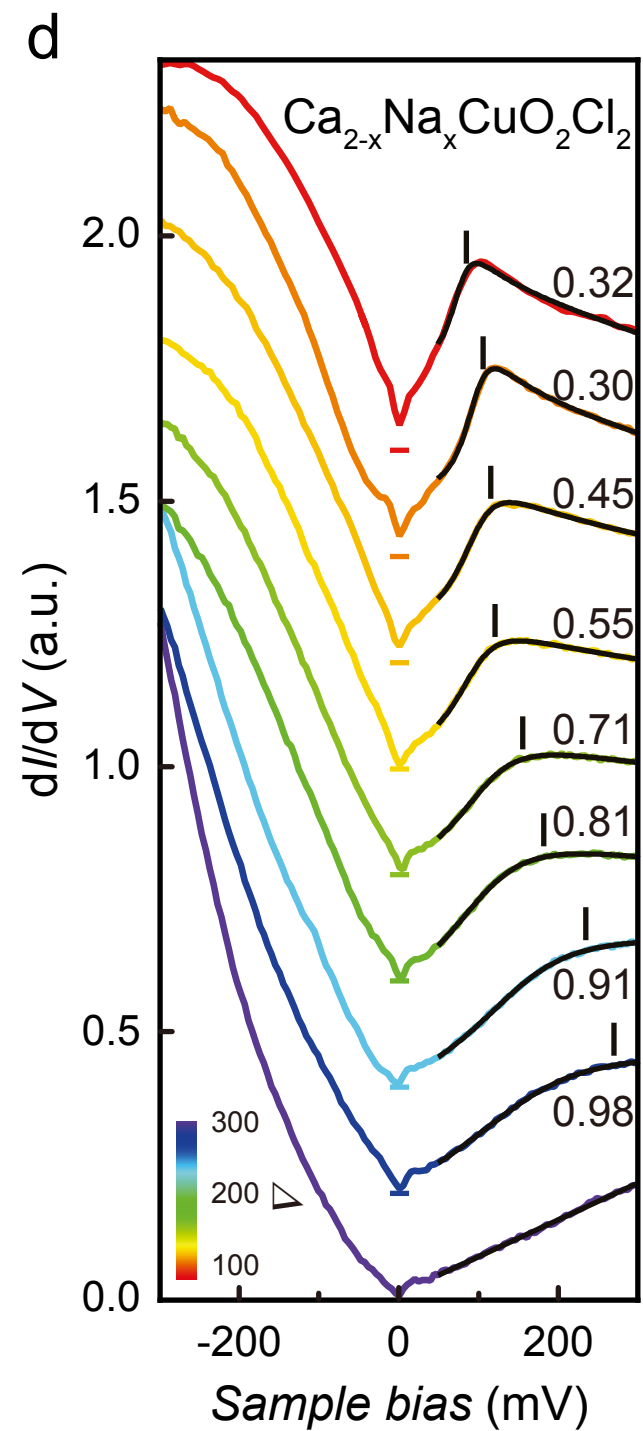

Figure 6

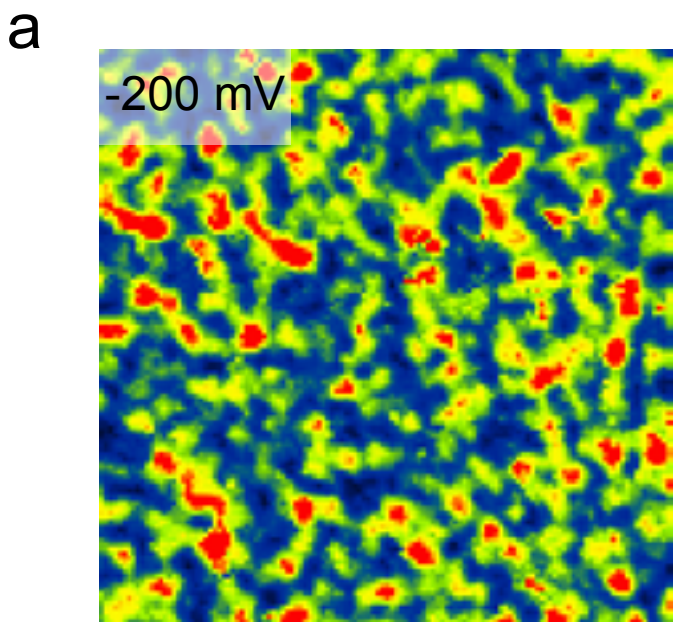
a
-200 mV

0 nS ▬▬ 0.20 nS

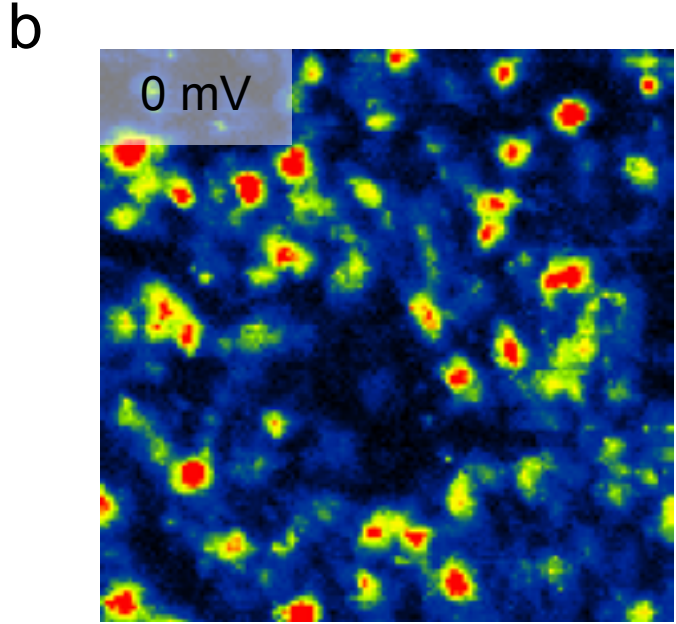
b
0 mV

0 nS ▬▬ 0.08 nS

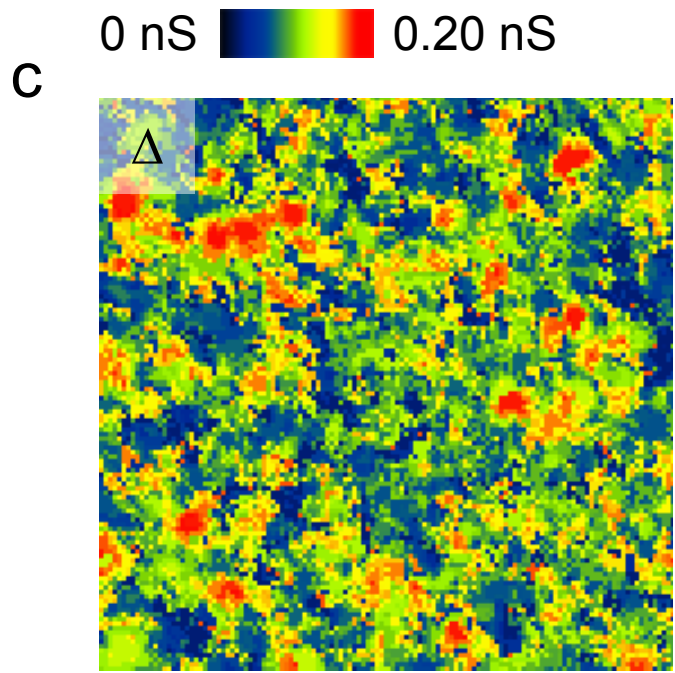
c
Δ

5 meV ▬▬ 40 meV

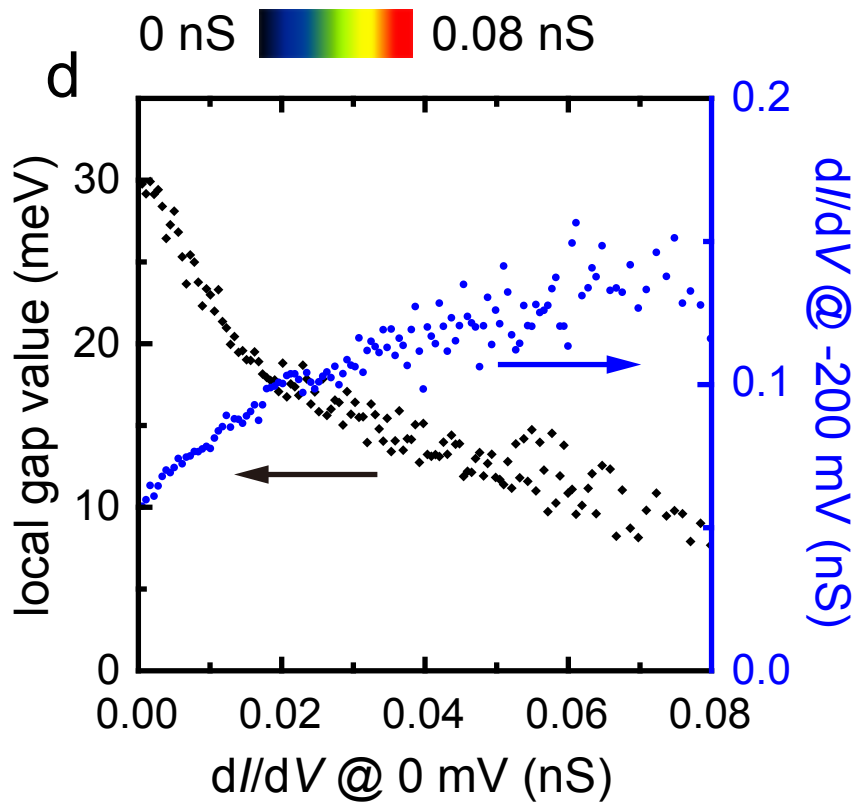